\newlength{\absize}
\documentclass[12pt]{article}

\usepackage{psfig}
\usepackage{epsfig}
\usepackage{epsf}
\usepackage{amssymb}

\renewcommand{\baselinestretch}{1.5}

\begin{document}
\thispagestyle{empty}
\pagestyle{empty}
\renewcommand{\thefootnote}{\fnsymbol{footnote}}
\newcommand{\starttext}{\newpage\normalsize
\pagestyle{plain}
\setlength{\baselineskip}{3ex}\par
\setcounter{footnote}{0}
\renewcommand{\thefootnote}{\arabic{footnote}}
}

\newcommand{\preprint}[1]{\begin{flushright}
\setlength{\baselineskip}{3ex}#1\end{flushright}}
\renewcommand{\title}[1]{\begin{center}\LARGE
#1\end{center}\par}
\renewcommand{\author}[1]{\vspace{2ex}{\Large\begin{center}
\setlength{\baselineskip}{3ex}#1\par\end{center}}}
\renewcommand{\thanks}[1]{\footnote{#1}}
\renewcommand{\abstract}[1]{\vspace{2ex}\normalsize\begin{center}
\centerline{\bf Abstract}\par\vspace{2ex}\parbox{\absize}{#1
\setlength{\baselineskip}{2.5ex}\par}
\end{center}}

\newlength{\eqnparsize}
\setlength{\eqnparsize}{.95\textwidth}

\setlength{\jot}{1.5ex}
\newcommand{\figsize}{\small}
\renewcommand{\bar}{\overline}
\font\fiverm=cmr5
\input prepictex
\input pictex
\input postpictex
\input{psfig.sty}
\newdimen\tdim
\tdim=\unitlength

\setcounter{bottomnumber}{2}
\setcounter{topnumber}{3}
\setcounter{totalnumber}{4}
\renewcommand{\bottomfraction}{1}
\renewcommand{\topfraction}{1}
\renewcommand{\textfraction}{0}

\def\draft{
\renewcommand{\label}[1]{{\quad[\sf ##1]}}
\renewcommand{\ref}[1]{{[\sf ##1]}}
\renewenvironment{thebibliography}{\section*{References}}{}
\renewcommand{\cite}[1]{{\sf[##1]}}
\renewcommand{\bibitem}[1]{\par\noindent{\sf[##1]}}
}

\def\theequation{\thesection.\arabic{equation}}
\preprint{\#HUTP-02/A050\\ 9/02}
\title{Quantum Moduli Spaces of 
Linear and Ring Mooses\thanks{
Research supported in part by the National Science Foundation under
grant number NSF-PHY/98-02709.
}}
\author{
Girma Hailu\thanks{hailu@feynman.harvard.edu}
\\
\small\sl Lyman Laboratory of Physics \\
\small\sl Harvard University \\
\small\sl Cambridge, MA 02138
}

\abstract{
Quantum moduli spaces of four dimensional $SU(2)^{r}$
linear and ring moose theories with $\mathcal{N}=1$ supersymmetry and
link chiral superfields in the fundamental representation are 
produced starting from simple pure gauge theories of disconnected 
nodes.
}

\starttext

\setcounter{equation}{0}
\section{{\label{sec:Intr}Introduction}}

Moose \cite{G-1} diagrams give succinct graphical representations of the transformations of 
matter fields under gauge (and global) symmetries. The gauge symmetries
are represented by nodes and the matter fields by links. The recent interest in moose theories 
is because a class of moose diagrams has been shown to transform
into a description of extra dimensions \cite{ACG-1,HPW-1}. Furthermore, the ``theory space''
of mooses has been used to build models and investigate various fundamental issues such as 
 electroweak symmetry breaking and accelerated
grand unification \cite{ACG-2}.
The transformation of a moose diagram into a description of 
extra dimensions occurs when the link 
fields develop vacuum expectation value (VEV) and ``hop'' between
the nodes. It is well known that supersymmetric gauge theories
have larger moduli spaces of vacua \cite{seiberg-1} than
non-supersymmetric gauge theories. Therefore, supersymmetric mooses could
provide richer theory spaces for model building. 

Our interest in this note is the construction of the quantum moduli spaces of 
$\mathcal{N}=1$ supersymmetric
$SU(2)^{r}$ linear and ring mooses where the gauge group
at each node is $SU(2)$ and the links are chiral superfields that
transform as fundamentals under the nearest gauge groups and as
singlets under the rest.
We will obtain nontrivial quantum moduli
spaces by starting from
pure gauge theories of disconnected nodes and exploiting simple and
efficient integrating in \cite{ILS-1,Intriligator-1} and out procedures. 
Explicit parameterization of the vacua in terms of gauge invariant objects
constructed out of the chiral superfields will be found.
A generic point in the moduli space of the ring moose has an unbroken $U(1)$
gauge symmetry and the ring moose is in the Coulomb phase.
We will find two singular submanifolds with modulus that is a nontrivial function 
of all the independent gauge invariants
needed to parameterize the moduli space of the ring moose. 
The Seiberg-Witten elliptic curve that describes the
quantum moduli space of the ring moose will follow from our computation.

Seiberg-Witten elliptic curves of the ring moose were computed in \cite{CEFS} using a different
method where it was started with the curve for a ring with two nodes given in  
 \cite{IS-1}  and  various asymptotic limits and symmetry arguments were used to obtain the curve for 
a ring moose with three nodes. The result was then 
generalized to the curve for a ring with arbitrary number of nodes. Here we will directly and 
explicitly compute the singularities of the quantum moduli space and the corresponding 
Seiberg-Witten elliptic curve for a ring moose with arbitrary number of nodes.
Our results agree with \cite{IS-1} for a ring with two nodes and with  \cite{CEFS} for a ring 
with three nodes.
We believe that the curve given in \cite{CEFS} is incorrect for ring mooses with four or more nodes. 

An expanded version of this note that also includes more topics and results is 
presented in \cite{H-1}. 

\setcounter{equation}{0}
\section{\label{sec-linear-su2r}Quantum moduli space of the linear moose\protect
\footnote{Quantum moduli
space constraint relations
for a linear moose with two and more nodes were first shown to us by Howard Georgi. Many results in
this section overlap with results in \cite{S-H}.
}}

Consider a four dimensional $\mathcal{N}=1$ supersymmetric  $SU(2)^{r}$ linear moose theory
with  matter content 
shown in Table \ref{lineartable1}. The equivalent moose diagram representation is 
shown in Figure \ref{fig-lmoose2}. 
An internal chiral superfield  
$Q_{i}$ links the $i^{th}$ and $(i+1)^{th}$ nodes. 
The internal
link $Q_{i}$ is a doublet that
transforms as $(\square,\,\square)$ under $SU(2)_{i}\times SU(2)_{i+1}$ and as
singlet under all the other gauge groups.  One of the external links $Q_{0}$ transforms as 
$\square$ under $SU(2)_{1}$ and the second external link $Q_{r}$ transforms as $\square$ under $SU(2)_{r}$.
Each external link is two doublets with $SU(2)$ subflavor symmetry.

\begin{table}[htb]
\begin{center}
\begin{tabular} {c|c c c c c|c c|}
\vspace{-0.25in}
&
\multicolumn{5}{|c|}{Gauge symmetries}
&
\multicolumn{2}{|c|}{Subflavor symmetries}
\\
&
$SU(2)_{1}$&
$SU(2)_{2}$&
$SU(2)_{3}$&
$\cdots $&
$SU(2)_{r}$&
$SU(2)_{1}$&
$SU(2)_{2}$
\\ \hline 
\vspace{-0.25in}
$Q_{0}$&
$\square $&
$1$&
$1$&
$\cdots $&
$1$&
$\square $&
$1$\\
\vspace{-0.25in}
$Q_{1}$&
$\square $&
$\square $&
$1$&
$\cdots $&
$1$&
$1$&
$1$\\ 
\vspace{-0.25in}
$Q_{2}$&
$1$&
$\square $&
$\square $&
$\cdots $&
$1$&
$1$&
$1$\\
\vspace{-0.25in}
$\vdots$ &
$\vdots$&
$\vdots$&
$\vdots$&
$\ddots$&
$\vdots$&
$\vdots$&
$\vdots$
\\
\vspace{-0.25in}
$Q_{r}$&
$1$&
$1$&
$1$&
$\cdots$&
$\square $&
$1$&
$\square $
\vspace{0.25in}\\
\hline
\end{tabular}
\end{center}
\caption{
Matter content of the linear moose.}
\label{lineartable1}
\end{table}

{\figsize\begin{figure}[htb]
{\centerline{\epsfxsize=15cm \epsfbox{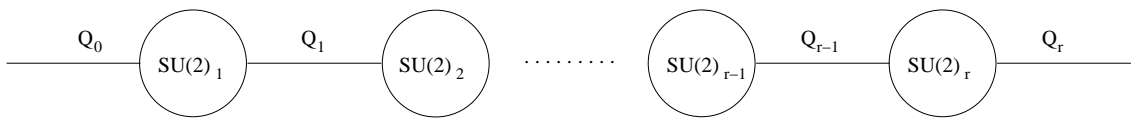}}}
\caption{\figsize\sf \label{fig-lmoose2}
Linear moose with $r$ nodes and $r+1$ links. The external links $Q_{0}$ and $Q_{r}$ each have
one color and one subflavor indices and each internal link has
two color indices.
}\end{figure}}

We will compute the quantum moduli space of this theory
starting from pure disconnected
gauge groups and integrating in the link fields. 
Gaugino condensation in the pure gauge theory gives
a nonperturbative superpotential,\begin{equation}
W_{\mathrm{d}}=\sum _{i=1}^{r}2\epsilon _{i}\Lambda _{0i}^{3},\label{sp-1}
\end{equation}
where each $\epsilon _{i}=\pm 1$ labels the two
vacua due to the breaking of the $Z_{4}$ $R$ symmetry to $Z_{2}$
 and $\Lambda _{0i}$ is the
nonperturbative dynamical scale of $SU(2)_{i}$.  
Our notation for the dynamical scales is $\Lambda
_{0i}$ for the scale of $SU(2)_{i}$ with no link,
$\Lambda _{id}$ when there is one link, and $\Lambda
_{i}$ when there are two links attached.
The scale $\Lambda _{i}$  is related to $\Lambda _{0i}$ by threshold matching of
the gauge coupling running at the masses $m_{i-1}$ and $m_{i}$ of
$Q_{i-1}$ and $Q_{i}$ respectively,
\begin{equation}
\Lambda _{0i}^{6}=\Lambda _{i}^{4}m_{i-1}m_{i}.\label{scales-1}
\end{equation}

There are $r+1$ link
chiral superfields each with four complex degrees of freedom. We can construct
 a total of $\frac{1}{2}(r^{2}+3r+8)$ gauge singlets given by
determinants of products of one to $r$ consecutive link superfields, and
the product of all the chiral superfields:
\begin{equation}
\mathrm{det }\,(Q_{i}),\label{lin-qi}
\end{equation}
\begin{equation}
\quad\mathrm{det }\,(Q_{i}Q_{i+1}),\quad \,\,\cdots,\,\,
\mathrm{det }\,(Q_{0}Q_{1}\cdots Q_{r-1}),\quad \mathrm{det }\,(Q_{1}Q_{2}\cdots Q_{r}),\label{lin-qis}
\end{equation}
\begin{equation}
\mathrm{ and} \quad
Q_{0}Q_{1}\cdots Q_{r}.\label{eq:linear-s1}\end{equation}

For a generic linear moose, the gauge symmetry is completely
broken and $3r$ of the complex degrees of freedom become massive or are eaten
by the super Higgs mechanism.
Consequently, there are only $4(r+1)-3r=r+4$ massless complex degrees
of freedom left. Because we have $\frac{1}{2}(r^{2}+3r+8)$ gauge
singlets, there must be $\frac{1}{2}(r^{2}+3r+8)-(r+4)=r(r+1)/2$
constraints. We claim that a constraint involving the determinants of only subsegments
of the moose chain given in (\ref{lin-qis}) are not modified by the extra links and nodes.
We can see that as follows: Consider the determinant of a subsegment,
$\mathrm{det}\,(Q_{i}Q_{i+1}\cdots Q_{j})$. The color indices from
the gauge groups $SU(2)_{i}$ and $SU(2)_{j}$ are not contracted
with the colors of $SU(2)_{i-1}$ and $SU(2)_{j+1}$ respectively.
Consequently, these adjoining gauge groups behave like global
subflavor symmetries. This amounts to saying that as far as
$\mathrm{det}\,(Q_{i}Q_{i+1}\cdots Q_{j})$ is concerned, the moose
chain is cut off at the $(i-1)^\mathrm{th}$ and $(j+1)^\mathrm{th}$ nodes.
Therefore, finding a constraint for
$\mathrm{det}\,(Q_{i}Q_{i+1}\cdots Q_{j})$ is not an independent
problem. Thus all the $r(r+1)/2$ moduli space constraints can be easily deduced from the one constraint 
which can be parameterized by the $r+5$ 
independent gauge singlets:
\begin{equation}
M_{i}\equiv\frac{1}{2}(Q_{i})_{\alpha _{i}\beta _{i}}(Q_{i})_{\alpha' _{i}\beta' _{i}}
\epsilon ^{\alpha _{i}\alpha' _{i}}\epsilon ^{\beta _{i}\beta' _{i}}=\mathrm{det}\,(Q_{i})\label{gaugeinv-1}
\end{equation}
and
\begin{equation}
T_{fg}\equiv \frac{1}{2}(Q_{0})_{f\beta _{0}}(Q_{1})_{\alpha _{1}\beta _{1}}(Q_{2})_{\alpha _{2}\beta _{2}}\cdots (Q_{r})_{\alpha _{r}g}\epsilon ^{\beta _{0}\alpha _{1}}\epsilon ^{\beta _{1}\alpha _{2}}\cdots \epsilon ^{\beta _{r-1}\alpha _{r}}.\label{gaugeinv-2}\end{equation}
where $\alpha_{i}$, $\beta_{i}$ are color indices and $f$, $g$ are subflavor indices.
For $M_{0}$ and $M_{r}$ one of the indices in $Q_{0}$ and $Q_{r}$ is for subflavor.

In order to integrate in the link fields to the pure gauge theory of disconnected nodes, 
first we use (\ref{scales-1}) to replace $\Lambda _{0i}^{3}\rightarrow 
(\Lambda _{i}^{4}m_{i-1}m_{i})^{\frac{1}{2}}$ in (\ref{sp-1}). We then write 
\begin{equation}
W=W_{\mathrm{d}}(\mathrm{with }\,\Lambda _{0i}^{3}\rightarrow 
(\Lambda _{i}^{4}m_{i-1}m_{i})^{\frac{1}{2}})+
W_{\mathrm{tree,d}}-W_{\mathrm{tree}}.\label{eq:inter-2a}
\end{equation}
$W_{\mathrm{tree}}$  is a tree level superpotential that contains couplings to all the independent gauge invariants,
\begin{equation}
W_{\mathrm{tree}}=\mathrm{tr}\,(c\,T{{{{}}}})+\sum _{i=0}^{r}m_{i}M_{i},
\label{eq:super-all-1aa}
 \end{equation}
where $c$ is a constant $2 \times 2$ matrix.
The term $W_{\mathrm{tree,d}}$ is computed by picking up any one link chiral superfield $Q_{k}$ and integrating  out 
$Q_{k}$ in the gauge and flavor invariant tree
level superpotential%
\footnote{This point is discussed in detail in \cite{H-1}.}%
$\,\mathrm{tr}\,(c\,T)+m_{k}M_{k}\label{super-nq-1}$. 
This gives \begin{equation} 
W_{\mathrm{tree,d}}=-\frac{\mathrm{det}\,(c)}{m_{k}}\,\mathrm{det(}Q_{0}Q_{1}\ldots Q_{k-1})
\,\mathrm{det(}Q_{k+1}Q_{k+2}\ldots Q_{r}).\label{eq:super-nq-2}
\end{equation}
We will see that the final result on the moduli space constraint does
not depend on $k$. 
For simplicity of notation, we introduce a more general way of representing
products of consecutive link chiral superfields and define
\begin{equation}
T_{(i,j)}\equiv Q_{i}Q_{i+1}\ldots Q_{j}.\label{eq:qprod-a}
\end{equation}
Note that $T_{(i,\,j)}$ is a $2 \times 2$ matrix with hidden indices.
The superpotential we need for integrating in all the independent gauge singlets
 is then, putting (\ref{sp-1}),  (\ref{eq:super-all-1aa}) 
and (\ref{eq:super-nq-2}) in (\ref{eq:inter-2a}),
\begin{eqnarray}
 W & =&  2\sum _{i=1}^{r}\epsilon _{i}(\Lambda _{i}^{4}m_{i-1}m_{i})^{\frac{1}{2}}-\frac{\mathrm{det}\,(c)}
{m_{k}}\,\mathrm{det}\,T_{(0,\, k-1)}\,\mathrm{det}\,T_{(k+1,\, r)}\nonumber \\
 &  & -\mathrm{tr}\,(c\,T_{(0,r)})-\sum _{i=0}^{r}m_{i}M_{i}.\label{eq:super-all-1}
\end{eqnarray}
The integrating in procedure is completed by minimizing (\ref{eq:super-all-1}) with the coupling constants 
  $m_{i}$ and $c$.  

Integrating out $m_{i}$ and $c$ in (\ref{eq:super-all-1}),
recursively solving for $m_{i}$ and $c$, and putting into (\ref{eq:super-all-1})
gives $W=0$ and a quantum moduli space constrained by
\begin{equation}
\mathrm{det}\,T_{(0,\, r)}-\frac{\mathrm{det}\,T_{(0,\, k-1)}\,\mathrm{det}\,T_{(k+1,\, r)}}{\Omega_{(0,\, k-1)}
\Omega_{(k+1,\, r)}}\Omega_{(0,\, r)}=0,\label{eq:moduli-a}\end{equation}
where we have introduced  $\Omega_{(i,\, j)}$ functions to simplify our
notation. The $\Omega$ functions are defined by
\begin{eqnarray}
\Omega_{(i,\, j)} & \equiv  & \prod _{q=i}^{j}M_{q}-\sum _{p=i+1\textrm{ }}^{j}
\Bigl (\Lambda _{p}^{4}\prod _{q\neq p-1,\, p}M_{q}\Bigr )+\sum _{p=i+1\textrm{ }}^{j-2}
\sum _{l=0}^{j-p-2}\Bigl (\Lambda _{p}^{4}\Lambda _{p+l+2}^{4}
\prod _{q\neq p-1,\, p,p+l+1,\, p+l+2}M_{q}\Bigr )\nonumber \\
 &  & -\cdots +(-1)^{(j-i+1)/2}\prod _{p=1}^{(j-i+1)/2}
\Lambda _{i+2p-1}^{4},\textrm{   }\label{eq:om-odd}
\end{eqnarray}
 if $j-i$ is odd, and
\begin{eqnarray}
\Omega_{(i,\, j)} & \equiv  & \prod _{q=i}^{j}M_{q}-\sum _{p=i+1\textrm{ }}^{j}
\Bigl (\Lambda _{p}^{4}\prod _{q\neq p-1,\, p}M_{q}\Bigr )+\sum _{p=i+1\textrm{ }}^{j-2}
\sum _{l=0}^{j-p-2}\Bigl (\Lambda _{p}^{4}\Lambda _{p+l+2}^{4}
\prod _{q\neq p-1,\, p,p+l+1,\, p+l+2}M_{q}\Bigr )\nonumber \\
 &  & -\cdots +(-1)^{(j-i)/2}\sum _{q=0\textrm{ }}^{(j-i)/2}\Bigl (M_{i+2q}
\prod _{p=0}^{q-1 \textrm{}}\Lambda _{i+2q-2p-1}^{4}\prod _{l=1}^{(j-i)/2-q}
\Lambda _{i+2q+2l}^{4}\Bigr ),\textrm{  }\label{eq:om-even}
\end{eqnarray}
if $j-i$ is even. We take $j>i$ unless explicitly stated. When $i=j$, we have $\Omega_{(i,\, i)}=\mathrm{det}M_{i}$.
Some recursion relations among the $\Omega$ functions are given in \cite{H-1}.

Thus the quantum moduli space is constrained by the recursion
relations given by (\ref{eq:moduli-a}). Note that $k$ in
(\ref{eq:moduli-a}) is arbitrary and could take any value from $0$
to $r$. As we have argued earlier in this section, a similar relation as
(\ref{eq:moduli-a}) should hold for a
subset of the linear chain, and we write a
more general form of the moduli space constraints as
\begin{equation}
\mathrm{det}\,{T_{(i,j)}}-\frac{\mathrm{det}\,{T_{(i,\, k-1)}}\,\mathrm{det}\,{T_{(k+1,\, j)}}}{
\Omega_{(i,\, k-1)}\Omega_{(k+1,\, j)}}\Omega_{(i,\, j)}=0.\label{eq:moduli-gen}
\end{equation}
Now we can easily prove that the result (\ref{eq:moduli-gen}) is independent
of $k$, since we can repeatedly use the same recursion relations to simplify the fractional factor in the
second term, 
and  (\ref{eq:moduli-gen}) gives
\begin{equation}
\mathrm{det}\,T_{(i,j)}-\Omega_{(i,\,j)}=0.\label{eq:moduli-simple}
\end{equation}
Note that (\ref{eq:moduli-simple}) gives $r(r+1)/2$
constraints that completely remove all the redundancy in the set
of gauge singlets.

The first few $\Omega $ functions are 
\begin{eqnarray}
\Omega_{(i,\, i+1)} & = & M_{i}M_{i+1}-\Lambda _{i+1}^{4},\label{eq:ap-omii1}\\
\Omega_{(i,\, i+2)} & = & M_{i}M_{i+1}M_{i+2}-\Lambda _{i+1}^{4}M_{i+2}-\Lambda _{i+2}^{4}M_{i},
\label{eq:ap-omii2}\\
\Omega_{(i,\, i+3)} & = & M_{i}M_{i+1}M_{i+2}M_{i+3}-\Lambda _{i+1}^{4}M_{i+2}M_{i+3}\nonumber \\
 &  & -\Lambda _{i+2}^{4}M_{i}M_{i+3}-\Lambda _{i+3}^{4}M_{i}M_{i+1}+\Lambda _{i+1}^{4}\Lambda _{i+3}^{4},\label{eq:ap-omii3}\\
\Omega_{(i,\, i+4)} & = & M_{i}M_{i+1}M_{i+2}M_{i+3}M_{i+4}-\Lambda _{i+1}^{4}M_{i+2}M_{i+3}M_{i+4}\nonumber \\
 &  &-\Lambda _{i+2}^{4}M_{i}M_{i+3}M_{i+4} -\Lambda _{i+3}^{4}M_{i}M_{i+1}M_{i+4}\nonumber \\
 &  &-\Lambda _{i+4}^{4}M_{i}M_{i+1}M_{i+2} +\Lambda _{i+1}^{4}\Lambda _{i+3}^{4}M_{i+4}\nonumber\\
 &  &+\Lambda _{i+1}^{4}\Lambda _{i+4}^{4}M_{i+2}+
\Lambda _{i+2}^{4}\Lambda _{i+4}^{4}M_{i}.\label{eq:ap-omii4}
\end{eqnarray}

\setcounter{equation}{0}
\section{\label{sec:ring}Quantum moduli space of the ring moose}

Now we can construct the quantum  moduli space of the ring moose starting from
the linear moose. 
The matter content of the ring moose is shown in Table \ref{ringtable1}.
\begin{table}[htb]
\begin{center}
\begin{tabular}{c|c c c c c c|} 
\vspace{-0.25in}
&
\multicolumn{6}{|c|}{Gauge symmetries}\\
&
$SU(2)_{1}$&
$SU(2)_{2}$&
$SU(2)_{3}$&
$\cdots $&
$SU(2)_{r-1}$&
$SU(2)_{r}$
\\
\hline
\vspace{-0.25in} 
$Q_{0}$&
$\square $&
$1$&
$1$&
$\cdots$&
$1$&
$\square$
\\
\vspace{-0.25in}
$Q_{1}$&
$\square $&
$\square $&
$1$&
$\cdots$&
$1$&
$1$
\\
\vspace{-0.25in} 
$Q_{2}$&
$1$&
$\square $&
$\square $&
$\cdots$&
$1$&
$1$
\\
\vspace{-0.25in}
$\vdots$&
$\vdots$&
$\vdots$&
$\vdots$&
$\ddots$&
$\vdots$&
$\vdots$
\\
\vspace{-0.25in}
$Q_{r-1}$&
$1$&
$1$&
$1$&
$\cdots$&
$\square $&
$\square $
\vspace{0.25in}\\
\hline
\end{tabular}
\end{center}
\label{ringtable1}
\caption{
Matter content of the ring moose.}
\end{table}
The equivalent moose diagram is shown in Figure \ref{figmring}.
{\figsize\begin{figure}[htb] 
{\centerline{\epsfxsize=6cm
\epsfbox{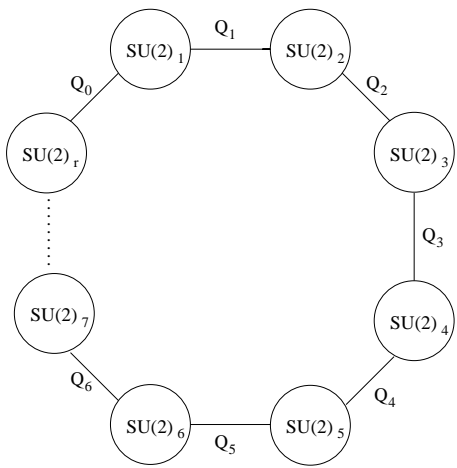}}} \caption{\figsize\sf\label{figmring} Ring
moose with $r$ nodes.  Each link has two color indices. }\end{figure}}
There are $r+1$ independent 
gauge singlets given by $M_{i}$ defined in (\ref{gaugeinv-1}), where $0 \le i \le r-1$ now, and 
\begin{equation}
U_{(0,\,r-1)}\equiv
 \frac{1}{2}(Q_{0})_{\alpha _{0}\beta _{0}}(Q_{1})_{\alpha _{1}\beta _{1}}
(Q_{2})_{\alpha _{2}\beta _{2}}\cdots (Q_{r-1})_{\alpha _{r-1}\beta _{r-1}}
\epsilon ^{\beta _{0}\alpha _{1}}\epsilon ^{\beta _{1}\alpha _{2}}\cdots
\epsilon ^{\beta _{r-1}\alpha _{0}}. \label{eq:eq:u0r}\end{equation}

We will start with the quantum moduli space constraint of the linear moose we found in 
Section \ref{sec-linear-su2r}. We will then
integrate out the external links.  Finally, a link field
that transforms as $(\square,\,\square)$ under $SU(2)_{r}\times
SU(2)_{1}$ will be integrated in to build the ring moose shown in Figure \ref{figmring}.
Since we can at the same time obtain the superpotential for a linear moose with only one external link,  
let us first integrate out $Q_{r}$. 
The superpotential for the linear moose without $Q_{r}$ is 
obtained by integrating out $M_{r}$, $T_{(0,\,r)}$ and $A$ in
\begin{equation}
W=A\Bigl (\mathrm{det}\,{T_{(0,\, r)}}-\Omega_{(0,\,r)}\Bigr
)+m_{r}M_{r}.\label{eq:winternalmin}
\end{equation}
The resulting superpotential is
\begin{equation}
W=\frac{\Lambda _{rd}^{5}\Omega_{(0,\, r-2)}}{\Omega_{(0,\, r-1)}},\label{eq:woneext}
\end{equation}
where $\Lambda _{rd}^{5}=\Lambda _{r}^{4}m_{r}$.
Next we integrate
out $Q_{0}$ by adding $m_{0}M_{0}$ to (\ref{eq:woneext}) and minimizing
with $M_{0}$ which  gives
the superpotential of the linear moose without the external links,
\begin{equation}
W=\frac{\Lambda _{1d}^{5}\Omega_{(2,\, r-1)}}{\Omega_{(1,\, r-1)}}+\frac{\Lambda _{rd}^{5}
\Omega_{(1,\, r-2)}}{\Omega_{(1,\, r-1)}}\pm 2\frac{(\Lambda _{1d}^{5}\Lambda _{rd}^{5}
\prod _{i=2}^{r-1}\Lambda _{i}^{4})^{1/2}}{\Omega_{(1,\, r-1)}}.\label{eq:winternal}
\end{equation}

The superpotential (\ref{eq:winternal})  can be interpreted
as follows: For the moose chain
with only internal links, the original $SU(2)^{r}$ gauge symmetry is completely broken and there is 
a new unbroken diagonal
$SU(2)_{D}$.  The first term comes from a single instanton in the broken
$SU(2)_{1}$ and infinite series of multi-instantons from the broken
$SU(2)_{2}$ to  $SU(2)_{r-2}$. Similarly, the second term comes
from
 a single instanton in the broken $SU(2)_{r}$ and an infinite series of multi-instantons from
the broken  $SU(2)_{2}$ to  $SU(2)_{r-2}$. These can be seen by using the explicit form of the
$\Omega$ functions and making an expansion of $\Omega_{(1,\, r-1)}^{-1}$ in powers of the
scales of $SU(2)_{2}$ to  $SU(2)_{r-2}$ .
 The last term comes from gaugino condensation
in the unbroken diagonal $SU(2)_{D}$. In fact, we can read off from  (\ref{eq:winternal}) that
the scale of the diagonal $SU(2)_{D}$ is
\begin{equation}
\Lambda_{D}=\Bigl(\frac{(\Lambda _{1d}^{5}\Lambda _{rd}^{5}
\prod _{i=2}^{r-1}\Lambda _{i}^{4})^{1/2}}{\Omega_{(1,\, r-1)}}\Bigr)^{\frac{1}{3}}.\label{lambda-all-1}
\end{equation}

Finally, we can construct the quantum moduli space of the ring moose
shown in Figure \ref{figmring} by integrating in $Q_{o}$.  
The new gauge singlets that appear in the ring moose which were not in the linear moose with 
only internal links are $M_{0}$ and  $U_{(0,\,r-1)}$ and the tree level superpotential we need
is 
\begin{equation}
W_{\mathrm{tree }}=b\,U_{(0,\,r-1)}+m_{0}M_{0},
\label{tree-ring-a}
\end{equation} 
where $b$ and $m_{0}$ are constants. The $W_{\mathrm{tree,d }}$ we need, because of 
the non-quadratic gauge singlet  $U_{(0,\,r-1)}$, is obtained by minimizing $b\,U_{(0,\,r-1)}+m_{0}M_{0}$ 
with $Q_{0}$ which gives 
$W_{\mathrm{tree,d }}=-\frac{b^{2}}{4m_{0}}\Omega_{(1,\, r-1)}$.
The
quantum moduli space constraint of the ring moose is then obtained
by minimizing
\begin{eqnarray}
W & = & \frac{m_{0}\Lambda _{1}^{4}\Omega_{(2,\, r-1)}}{\Omega_{(1,\, r-1)}}+
\frac{m_{0}\Lambda _{r}^{4}\Omega_{(1,\, r-2)}}{\Omega_{(1,\, r-1)}}\pm 2\frac{m_{0}(
\prod _{i=1}^{r}\Lambda _{i}^{4})^{1/2}}{\Omega_{(1,\, r-1)}}\nonumber \\
 &  & -\frac{b^{2}}{4m_{0}}\Omega_{(1,\, r-1)}-m_{0}M_{0}-b\,U_{(0,\,r-1)}\label{eq:wminring}
\end{eqnarray}
with $m_{0}$ and $b$ which gives $W=0$ and \begin{equation}
U_{(0,\,r-1)}^{2}+\Lambda _{1}^{4}\Omega_{(2,\, r-1)}+\Lambda _{r}^{4}\Omega_{(1,\, r-2)}-M_{0}
\Omega_{(1,\, r-1)}\pm 2(\prod _{i=1}^{r}\Lambda _{i}^{4})^{1/2}=0.\label{eq:moduli-ring-b}
\end{equation}
This is symmetric in all links and scales. 

Before we interpret (\ref{eq:moduli-ring-b}), let us first recall the Seiberg-Witten
hypothesis on the elliptic curve of an $SU(2)$ gauge theory.
According to the
Seiberg-Witten hypothesis \cite{SW-1}, the quantum moduli space of
an $SU(2)$ gauge theory coincides with the moduli space of the
elliptic curve $y^{2}=(x^{2}-u)^{2}-\Lambda ^{4}$, where $u$ is a gauge invariant coordinate 
and $\Lambda$ is the dynamical scale of the theory. 
The singularities of this curve are given by the zeros of the
discriminant $\Delta _{\Lambda }=(u^{2}-\Lambda ^{4})(2\Lambda)
^{8}$. This occurs at $u=\pm \Lambda ^{2}$ and $u=\infty $. The
first two singularities at $u=\pm \Lambda ^{2}$ are in the strong
coupling region, and there is a massless monopole at one and a
massless dyon at the other of these singularities. The singularity
at $u=\infty $ is in the semi-classical region.

Now let us rewrite (\ref{eq:moduli-ring-b}) as
\begin{equation}
u_{r} = \pm \Lambda_{(1,\,r)}^{2},\label{eq:u-lam-1}
\end{equation}
where
\begin{equation}
u_{r} \equiv U_{(0,\,r-1)}^{2}+\Lambda _{1}^{4}\Omega_{(2,\, r-1)}+\Lambda
_{r}^{4}\Omega_{(1,\, r-2)}-M_{0} \Omega_{(1,\, r-1)}\label{eq:ufun-lam}
\end{equation}
and
\begin{equation}
\Lambda_{(1,\,r)}^{2}
\equiv 2(\prod _{i=1}^{r}\Lambda _{i}^{4})^{1/2}\label{eq:ufun-lam2}
\end{equation}

Note that the modulus $u_{r}$ contains all the independent gauge invariants we needed to parameterize
the moduli space of the ring. What (\ref{eq:u-lam-1}) is telling us is that the
function $u_{r}$ is locked at $\pm \Lambda_{(1,\,r)}^{2}$.
In other words, (\ref{eq:u-lam-1}) gives two $r$ - complex dimensional singular submanifolds in the $r+1$ - complex 
dimensional moduli space spanned by all the independent gauge invariants. 
Giving large VEVs to the link fields breaks the original $SU(2)^{r}$
gauge symmetry into a diagonal $SU(2)_{D}$ with matter in the adjoint representation. 
The two singularities given by (\ref{eq:u-lam-1}) on the $u_{r}$ plane 
can be nothing but the two singularities in the strong coupling region of the $SU(2)_{D}$ gauge theory with 
$\mathcal{N}=2$ supersymmetry. 
The monodromies around these singularities on the $u_{r}$ plane must be the same as in 
Seiberg-Witten and 
the charge at
the singularity  $u_{r} = + \Lambda_{(1,\,r)}^{2}$
is that of a monopole and the charge at $u_{r} = - \Lambda_{(1,\,r)}^{2}$ is that of
a dyon. A generic point in the moduli space of 
the ring moose has unbroken $U(1)$ gauge symmetry and the ring moose is 
in the Coulomb phase.
Having obtained these singularities and because the $U(1)$ coupling coefficient is holomorphic, we have 
determined the 
elliptic curve that
parameterizes the Coulomb phase of the ring moose.  
Thus the quantum moduli space of the
ring moose can be parameterized by the elliptic curve
\begin{equation}
y^{2}=\Bigl (x^{2}-[U_{(0,\,r-1)}^{2}+\Lambda _{1}^{4}\Omega_{(2,\, r-1)}+\Lambda _{r}^{4}
\Omega_{(1,\, r-2)}-M_{0}\Omega_{(1,\, r-1)}]\Bigr )^{2}-4\prod _{i=1}^{r}\Lambda _{i}^{4}.
\label{eq:ssw-curve-a}\end{equation}

Note that although the singularities look the same as in Seiberg-Witten on the $u_{r}$ plane, they are 
$r$ - complex dimensional submanifolds with very nontrivial modulus given by (\ref{eq:ufun-lam}).   
Using the definition (\ref{eq:ufun-lam}) for
$u_{r}$ and the $\Omega $ functions given in
Section \ref{sec-linear-su2r}, the first few $u$ functions are
\begin{eqnarray}
u_{2} & = & U_{(0,\,1)}^{2}+\Lambda _{1}^{4}+\Lambda _{2}^{4}-M_{0}M_{1},\label{eq:ap4-u2}\\
u_{3} & = & U_{(0,\,2)}^{2}+\Lambda _{1}^{4}M_{2}+\Lambda _{2}^{4}M_{0}+\Lambda _{3}^{4}M_{1}-M_{0}M_{1}M_{2},
\label{eq:ap4-u3}\\
u_{4} & = & U_{(0,\,3)}^{2}+\Lambda _{1}^{4}M_{2}M_{3}+\Lambda _{2}^{4}M_{0}M_{3}+\Lambda _{3}^{4}M_{0}M_{1}\nonumber \\
 &  &+
\Lambda _{4}^{4}M_{1}M_{2}-\Lambda _{1}^{4}\Lambda _{3}^{4}-\Lambda _{2}^{4}\Lambda _{4}^{4}-M_{0}M_{1}M_{2}M_{3},
\label{eq:ap4-u4}\\
u_{5} & = & U_{(0,\,4)}^{2}+\Lambda _{1}^{4}M_{2}M_{3}M_{4}+\Lambda _{2}^{4}M_{0}M_{3}M_{4}+
\Lambda _{3}^{4}M_{0}M_{1}M_{4}\nonumber \\
 &  & +\Lambda _{4}^{4}M_{0}M_{1}M_{2}+\Lambda _{5}^{4}M_{1}M_{2}M_{3}-\Lambda _{1}^{4}\Lambda _{3}^{4}M_{4}-
\Lambda _{1}^{4}\Lambda _{4}^{4}M_{2}\nonumber \\
 &  &-\Lambda _{2}^{4}\Lambda _{5}^{4}M_{3}-
\Lambda _{3}^{4}\Lambda _{5}^{4}M_{1}-\Lambda _{2}^{4}\Lambda _{4}^{4}M_{0}-M_{0}M_{1}M_{2}M_{3}M_{4}.
\label{eq:ap4-u5}
\end{eqnarray}

Seiberg-Witten curves for the ring moose were computed in \cite{CEFS} using a different method. 
A method used in \cite{IS-1} to obtain the curve for the $r=2$ ring was continued in \cite{CEFS} 
to compute the curve for $r=3$. 
The idea was as follows:
Because giving large VEVs to the link fields breaks the
$SU(2)^{r}$ gauge symmetry into a diagonal
$SU(2)_{D}$ with matter in the adjoint representation, the theory
in effect becomes that of a single $SU(2)$ with $\mathcal{N}=2$
supersymmetry. The curve for $r=3$ was obtained by taking various asymptotic limits of 
the gauge singlet fields and the 
nonperturbative scales, comparing with
the $\mathcal{N}=2$ $SU(2)$ curve and imposing symmetries. 
The result for $r=3$ was then generalized
to the curve for a ring moose with arbitrary $r$.
Our results agree  with
\cite{IS-1} for $r=2$ and with \cite{CEFS} for $r=3$. However, we
do not agree with the curves in \cite{CEFS} for $r\geq 4$. Only few
terms in $u_{r}$ were obtained in \cite{CEFS}, which would give incorrect singular submanifolds in
moduli space. We are not suggesting that the method used in \cite{CEFS} is incorrect.
Here we have obtained the quantum moduli space directly by integrating in all
the independent link fields starting from a pure gauge theory of disconnected nodes and building the
ring moose via the linear moose. This is done for a ring with arbitrary
number of nodes without any need of imposing symmetries in the
nodes or links and without taking asymptotic limits; and the result is
automatically symmetric in all nodes and links.

\setcounter{equation}{0}
\section{Summary\label{summary}}

We have produced 
 nontrivial quantum moduli spaces  for $\mathcal{N}=1$ supersymmetric $SU(2)^{r}$
 linear and ring mooses 
starting from simple pure gauge theories of disconnected nodes
by integrating
in all matter link fields.  
For the ring moose, we obtained two singular submanifolds with modulus that is 
a function of all the independent 
gauge singlets we needed to parameterize the quantum moduli space.  
The Seiberg-Witten elliptic
curve that describes the quantum moduli space of the ring moose
followed from our computation. More details and results are 
presented in \cite{H-1}. 

\section*{Acknowledgements}

I am deeply grateful to Howard Georgi for various useful suggestions, comments and discussions.
I am also very thankful to him for sharing his original quantum moduli space relations of the
linear moose with me which initiated this work.  This research is supported in part by
the National Science Foundation under grant number NSF-PHY/98-02709.


\begin{thebibliography}{10}

\bibitem{G-1}H. Georgi, {}``A tool kit for builders of composite models,'' \emph{Nucl.
Phys}. \textbf{B266} (1986) 274.
\bibitem{ACG-1}N. Arkani-Hamed, A. G. Cohen and H. Georgi, {}`` (De)constructing
dimensions,'' \emph{Phys. Rev. Lett}., \textbf{86}, 4757(2001), hep-th/0104005.
\bibitem{HPW-1}C. T. Hill, S. Pokorski, and J. Wang, {}``Gauge invariant effective
Lagrangians for Kaluza-Klein modes,'' \emph{Phys. Rev.} \textbf{D64} (2001) 105005, hep-th/0104035.
\bibitem{ACG-2}N. Arkani-Hamed, A. G. Cohen and H. Georgi, {}``Electroweak symmetry
breaking from dimensional deconstruction,'' \emph{Phys. Lett}., \textbf{B513},
232(2001), hep-ph/0105239; ``Accelerated unification,''
hep-ph/0108089.
\bibitem{seiberg-1}N. Seiberg, {}``Exact results on the space of vacua of four dimensional
susy gauge theories,'' \emph{Phys. Rev.} \textbf{D49} (1994) 6857,
hep-th/9402004.
\bibitem{ILS-1}K. Intriligator, R. G. Leigh and N. Seiberg, {}``Exact superpotentials
in four dimensions,'' \emph{Phys. Rev.} \textbf{D50} (1994) 1092,
hep-th/9403198.
\bibitem{Intriligator-1}K. Intriligator, {}``''Integrating in'' and exact superpotentials
in 4d,'' \emph{Phys. Lett.} \textbf{b336} (1994) 409, hep-th/9407106.
\bibitem{CEFS}C. Csaki, J. Erlich, D. Freedman and W. Skiba, {}``$\mathcal{N}=1$
supersymmetric product group theories in the coulomb phase,'' \emph{Phys.
Rev.} \textbf{D56} (1997) 5209, hep-th/9704067.
\bibitem{IS-1}K. Intriligator and N. Seiberg, {}``Phases of $\mathcal{N}=1$ supersymmetric
gauge theories in four dimensions,'' \emph{Nucl. Phys}. \textbf{B431}
(1994) 551, hep-th/9408155.
\bibitem{H-1}G. Hailu, {}`'$\mathcal{N}=1$ supersymmetric $SU(2)^{r}$ moose theories,'' hep-th/0209266.
\bibitem{S-H}S. Chang and H. Georgi, ``Quantum modified mooses,'' hep-th/0209038.
\bibitem{SW-1}N. Seiberg and E. Witten, \emph{{}``}Monopole condensation and confinement
in $\mathcal{N}=2$ supersymmetric Yang-Mills theory,'' \emph{Nucl.
Phys}. \textbf{B426} (1994) 19, hep-th/9407081; \emph{{}``}Monopoles,
duality and chiral symmetry breaking in $\mathcal{N}=2$ supersymmetric
QCD,'' \emph{Nucl. Phys}. \textbf{B431} (1994) 484, hep-th/9408099.
\end{thebibliography}
\end{document}